\documentclass[a4paper]{PoS}
\usepackage{mcite}
\usepackage{graphicx}  
\usepackage{amsmath}   
\usepackage{amssymb}   
\usepackage[caption=false]{subfig}

\title{Differential cross section for the Higgs boson production
        in 4-lepton channel and $k_T$-factorization}

\ShortTitle{The Higgs boson production and $k_T$-factorization}

\author{\speaker{Vaibhav S. Rawoot}\\
        The Institute of Mathematical Sciences,\\
             IV Cross Road, CIT Campus, Chennai 600 113, India.\\
E-mail: \email{vaibhavrawoot@gmail.com}}

\author{Rashidul Islam\\
        The Institute of Mathematical Sciences,\\
             IV Cross Road, CIT Campus, Chennai 600 113, India.\\
        E-mail: \email{rislam@iitg.ernet.in}}

\author{Mukesh Kumar\\
        National Institute for Theoretical Physics,\\
                School of Physics and Mandelstam Institute for Theoretical Physics,\\
                University of the Witwatersrand, Johannesburg, Wits 2050, South Africa.\\
        E-mail: \email{mukesh.kumar@cern.ch}}

\abstract{
We present our study of the differential cross section for the Higgs boson production
in $k_T$-factorization framework. The $k_T$-factorization formalism includes a
convolution over unintegrated parton distribution functions (uPDF) and off-shell
parton level matrix element. The off-shell matrix element calculated considering
initial gluons to be off-shell. We have considered only gluon fusion process which
is dominant production mechanism for the Higgs boson production at LHC. We have used
Ciafaloni-Catani-Fiorani-Marchesini (CCFM) uPDF based on CCFM evolution equations.
We have compared our results with fixed order estimates up to NNLO+NNLL obtained
using HRes tool within collinear factorization framework as well as with the ATLAS
and CMS measurements of the corresponding differential distributions. This study
will play an important role in understanding differential cross section within
$k_T$-factorization framework.}

\FullConference{QCD Evolution 2016\\
		May 30-June 03, 2016\\
		National Institute for Subatomic Physics (Nikhef), Amsterdam}

\def\nl{\nonumber \\}

\newcommand{\bfk}{{ \ensuremath{\bf k} }}
\newcommand{\bkt}{{ \ensuremath{\bf k_\perp} }}
\newcommand{\kt}{k_\perp}
\newcommand{\bfp}{{ \ensuremath{\bf p} }}

\begin{document}



\section{Introduction}
\label{sec:intro}

The discovery of a $125$~GeV candidate of the Standard Model (SM) Higgs boson by
the ATLAS and CMS experiments~\cite{Aad:2012tfa,*Chatrchyan:2012xdj,
Khachatryan:2014jba,*Aad:2015zhl} opened a new plethora of studies. It has been
established beyond doubt that the new particle is a spin-0 Higgs boson rather than
a spin-2 particle~\cite{Chatrchyan:2012jja,*Aad:2013xqa}. In the minimal SM, the
electroweak symmetry breaking is achieved via the Higgs mechanism through a single
complex scalar doublet which leads to only one neutral physical scalar, $h$.
Non-minimal models assume the existence of additional charged
and neutral scalar Higgs particles. Hence on one hand we are concerned about its
couplings to fermions and gauge bosons and the results seems to confirm consistency
with the expected SM values so far~\cite{Khachatryan:2014jba,*Khachatryan:2014kca,
*Aad:2015mxa,*Aad:2015gba}. On the other hand it opens the possibility to probe
the production channel of the said Higgs boson. Here we are concerned with the
latter aspect of the Higgs boson studies.

At the LHC conditions, the gluon-gluon fusion channel, $gg \to h$ is the most dominant
one for the inclusive Higgs boson production~\cite{Wilczek:1977zn,*Georgi:1977gs,
*Ellis:1979jy,*Rizzo:1979mf,*Graudenz:1992pv,*Spira:1995rr}. Clearly, the gluon-gluon
fusion to the Higgs production is strongly dependent on the gluon density $xf_g(x,
\mu^2_F)$ in a proton which are usually described as a function of the Bjorken
variable, $x$ and hard scale, $\mu^2_F$ within the framework of the DGLAP evolution
equation~\cite{Gribov:1972ri,*Lipatov:1974qm,*Altarelli:1977zs,*Dokshitzer:1977sg}.
The cross sections are calculated from the hard matrix elements convoluted with gluon
density functions. This is the so called {\em collinear factorization} approach.
However, at the LHC energies, it is more appropriate to use the parton densities
which are explicit functions of it's transverse momentum, $k_T$~\cite{Andersson:2002cf,
*Andersen:2003xj,*Andersen:2006pg} (the so called unintegrated, i.e., $k_T$ dependent
parton distribution functions or uPDFs in short). These parton densities are described
by the BFKL evolution equation~\cite{Kuraev:1976ge,*Kuraev:1977fs,*Balitsky:1978ic} for
very small $x$ or the CCFM evolution equation~\cite{Ciafaloni:1987ur,*Catani:1989yc,
*Catani:1989sg,*Marchesini:1994wr} which is valid for both small and large $x$\footnote{
  CCFM evolution is equivalent to BFKL evolution in the limit of very small $x$, whereas
  similar to the DGLAP evolution for large $x$ and high $\mu^2_F$}.
As in the case of the collinear factorization, it is also possible to factorize the cross
section into a convolution of the hard matrix elements and gluon uPDFs. However, in
this case the matrix elements have to be taken off-shell and the convolution should
also be over $k_T$. This generalized factorization is called {\em $k_T$-factorization}
\cite{Gribov:1984tu,*Levin:1991ry,*Catani:1990eg,*Collins:1991ty}.

Recently the ATLAS and CMS collaborations at the LHC presented a measurement of
fiducial differential cross section of the Higgs boson in the four-lepton decay
channel~\cite{Aad:2014tca,Khachatryan:2015yvw}. In particular, they presented a
measurement of differential cross section in transverse momentum and rapidity of
the Higgs boson decay into four-leptons. The measurements seems to be well in
agreement with the theoretical calculation based on the collinear factorization
approach with fixed order calculation up to the next to next leading order (NNLO)
and including soft gluon resummation at small transverse momenta up to the next
to next leading logarithm (NNLL). The $k_T$-factorization approach where the effect
of intrinsic transverse momentum is taken into account has been a recent interest
in the case of the Higgs boson production~\cite{Lipatov:2005at,*Ryskin:1999yq}.
The $k_T$-factorization approach provides an advantage to understand an intrinsic
dynamic of partons usually formulated in the form of uPDFs~\cite{Lipatov:2014mja}.

Here we have presented a study for the differential cross section of the Higgs boson
in the four-lepton channel using $k_T$-factorization approach. We have used the CCFM
evolution equations for the uPDFs. We have evaluated off-shell matrix element for the
partonic subprocess $g^*g^*\to h\to ZZ\to 4\ell, \ell=e,\mu$. We presented our results
for the differential cross section of the Higgs boson production in the four-lepton
channel in the framework of $k_T$-factorization.


\section{Formalism}

To estimate differential cross section for the Higgs boson production in four-lepton
decay channel, we have calculated off-shell matrix element for process $g^*g^*\to h
\to ZZ\to 4\ell$. We have considered effective field theory approach in our calculation.
Effective Lagrangian for the gluon coupling to the Higgs boson is
\begin{equation}
 {\cal L}_{ggh}
 =
 \frac{\alpha_S}{12 \pi} (\sqrt{2} G_F)^{1/2} G^a_{\mu\nu} G^{a\mu\nu} h,
 \label{lag-ggh}
\end{equation}
where $\alpha_s$ and $G_F$ are the strong coupling constant and Fermi coupling constant
respectively. $G^a_{\mu\nu}$ is the gluon field strength tensor and $h$ is the Higgs
scalar field.

Triangle vertex for $Hgg$ effective coupling in the infinite top mass limit is
\begin{equation}
  T^{\mu \nu,\, ab}_{ggH}(k_1,k_2) = i \delta^{ab} {\frac{\alpha_s}{3\pi}} 
  (\sqrt{2} G_F)^{1/2} \left[k_2^\mu k_1^\nu - (k_1 \cdot k_2) g^{\mu \nu}\right].
  \label{ver-ggh}
\end{equation}
Using Eq.~\eqref{ver-ggh} and considering transverse momentum of initial gluon to be
non-zero $k^2_1 = - \bfk^2_{\perp 1} \neq 0$ and $k^2_2 = - \bfk^2_{\perp 2} \neq 0$
we derived matrix element for process $g^*(k_1)g^*(k_2)\to h\to ZZ\to \ell(p_1)\bar
\ell(p_2)$ $\ell'(p_3)\bar{\ell}'(p_4)$. Matrix element we have obtained is
\begin{multline}
 |\mathcal{M}|^2
 = \frac{2}{9}\frac{\alpha^2_s}{\pi^2} \frac{\,m^4_Z}{v^4}
  \frac{\left[({\bkt}_1+{\bkt}_2)^2+{\hat s}\right]^2}{(\hat{s}-m^2_h)^2+\Gamma^2_h m^2_h}
  \cos^2\phi \times
 \\
  \frac{[(p_1\cdot p_4)(p_2\cdot p_3)\{2g^2_Lg^2_R\}+(p_1\cdot p_3)(p_2\cdot p_4)\{g^4_L+g^4_R\}]}
  {[(2 p_1\cdot p_2-m^2_Z)^2+\Gamma^2_Z m^2_Z][(2 p_3\cdot p_4-m^2_Z)^2+\Gamma^2_Z m^2_Z]},  
  \label{offshellm}
\end{multline}
and
\begin{align}
 g_L = \frac{g_W}{\cos\theta_W} \Big( - \frac{1}{2} + \sin^2\theta_W \Big),
 \qquad
 g_R = \frac{g_W}{\cos\theta_W} \sin^2\theta_W, \quad \text{and} \quad
 v=(\sqrt{2}G_F)^{-1/2},
\end{align}
where $\Gamma_h$ and $\Gamma_Z$ are total decay width of the Higgs boson and $Z$ boson
respectively. ${\kt}_1$ and ${\kt}_2$ are intrinsic transverse momentum of the initial
gluons. $\phi$ is the azimuthal angle between transverse momentum of the initial gluons
${\bkt}_1$ and ${\bkt}_2$. $m_Z$ and $m_h$ are the Higgs boson and Z boson masses
respectively. Partonic center of mass energy is denoted by $\hat{s}$. $\theta_W$ and
$g_W$ are weak mixing angle and coupling of weak interaction respectively.

The summation over polarization of off-shell initial gluon considered according to
$k_T$-factori\-zation prescription is
\begin{equation}
 \sum \epsilon^\nu \epsilon^{*\mu} = \frac{\bkt^\mu \bkt^\nu}{\bfk^2_{\perp}}.
\end{equation}

Within the framework of $k_T$-factorization, cross section for the process $p(P_1)+p(P_2)
\to h\to ZZ\to \ell(p_1)\bar{\ell}(p_2)\ell'(p_3)\bar{\ell}'(p_4)$ is written as
\begin{align}
 {\sigma}
 =&
 \int \frac{dx_1}{x_1}\frac{dx_2}{x_2} \frac{d^2\bfk_{\perp 1}}{\pi} \frac{d^2\bfk_{\perp 2}}{\pi}
 \frac{1}{x_1x_2s (2^{12})(\pi)^8}
 \frac{d^3p_{1}}{2p_{10}} \frac{d^3p_{2}}{2p_{20}} \frac{d^3p_{3}}{2p_{30}}\frac{d^3p_{4}}{2p_{40}}
 \nl
 &\times
 |\bar{\mathcal{M}}|^2 \delta^4(k_1+k_2-p_1-p_2-p_3-p_4)
 f_g(x_1,\bfk^2_{\perp 1})f_g(x_2,\bfk^2_{\perp 2}).
\label{facttmd1}
\end{align}
Simplifying Eq. \eqref{facttmd1} we obtained our final expression for cross section as
\begin{align}
 {\sigma}
 =&
 \int dy_1 dy_2 dy_3 dy_4 d\bfp^2_{1T} d\bfp^2_{2T} d\bfp^2_{3T} 
 {d\bfk^2_{\perp 1}}{d\bfk^2_{\perp 2}}
 \frac{d\phi_1}{2\pi}\frac{d\phi_2}{2\pi}
 \nl
 &\times 
 \frac{1}{(2^{12})\pi^5 (x_1 x_2 s)^2}
 |\bar{\mathcal{M}}|^2
 f_g(x_1,\bfk^2_{\perp 1})f_g(x_2,\bfk^2_{\perp 2}),
\label{facttmd5}
\end{align}
with longitudinal momentum fraction $x_1$ and $x_2$ of initial gluons are 
\begin{equation}
 x_1=\frac{|\bfp_{1T}|}{\sqrt{s}} e^{y_1}+\frac{|\bfp_{2T}|}{\sqrt{s}}e^{y_2}
 +\frac{|\bfp_{3T}|}{\sqrt{s}} e^{y_3}+\frac{|\bfp_{4T}|}{\sqrt{s}} e^{y_4},
\end{equation}
\begin{equation}
 x_2=\frac{|\bfp_{1T}|}{\sqrt{s}} e^{-y_1}+\frac{|\bfp_{2T}|}{\sqrt{s}} e^{-y_2}
 +\frac{|\bfp_{3T}|}{\sqrt{s}} e^{-y_3}+\frac{|\bfp_{4T}|}{\sqrt{s}} e^{-y_4},
\end{equation}
and 
\begin{equation}
\bfk_{\perp 1}+\bfk_{\perp 2}=\bfp_{1T}+{\bfp_{2T}}+\bfp_{3T}+\bfp_{4T},
\end{equation}
where $\phi_1$ and $\phi_2$ are the azimuthal angle of ${\bkt}_1$ and ${\bkt}_2$
respectively. $y$ and $p_{T}$ are rapidity and transverse momentum of the final
state leptons. Hadronic center of mass energy is denoted by $s$. 

We have calculated cross section for the Higgs boson decay as a function of rapidity
of four-leptons ($y$) and transverse momentum of the four-leptons ($p_T$). 
We have used Eq. \eqref{facttmd5} together with off-shell matrix element
given in Eq. \eqref{offshellm} for process $g^*g^*\to h\to ZZ\to 4\ell$, $\ell=e, \mu$, to calculate differential cross section.

\begin{figure}[!ht]
  \centering
  \subfloat[]{\includegraphics[trim=0 0 0 0,clip,height=0.44\linewidth,width=0.49\linewidth]{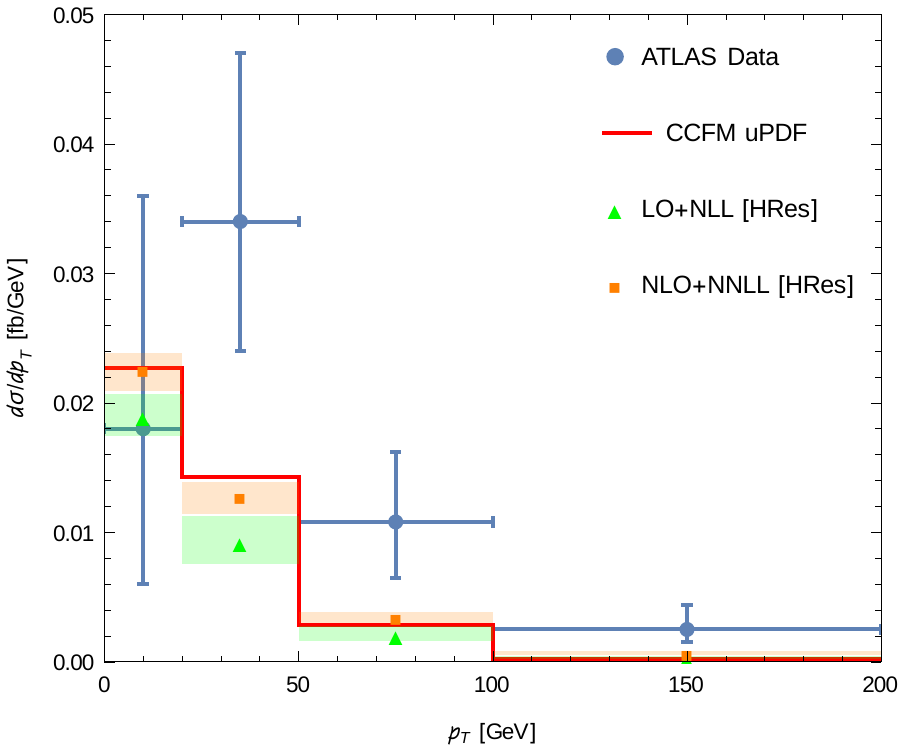}\label{fig1a}}
  \subfloat[]{\includegraphics[trim=0 0 0 0,clip,height=0.44\linewidth,width=0.49\linewidth]{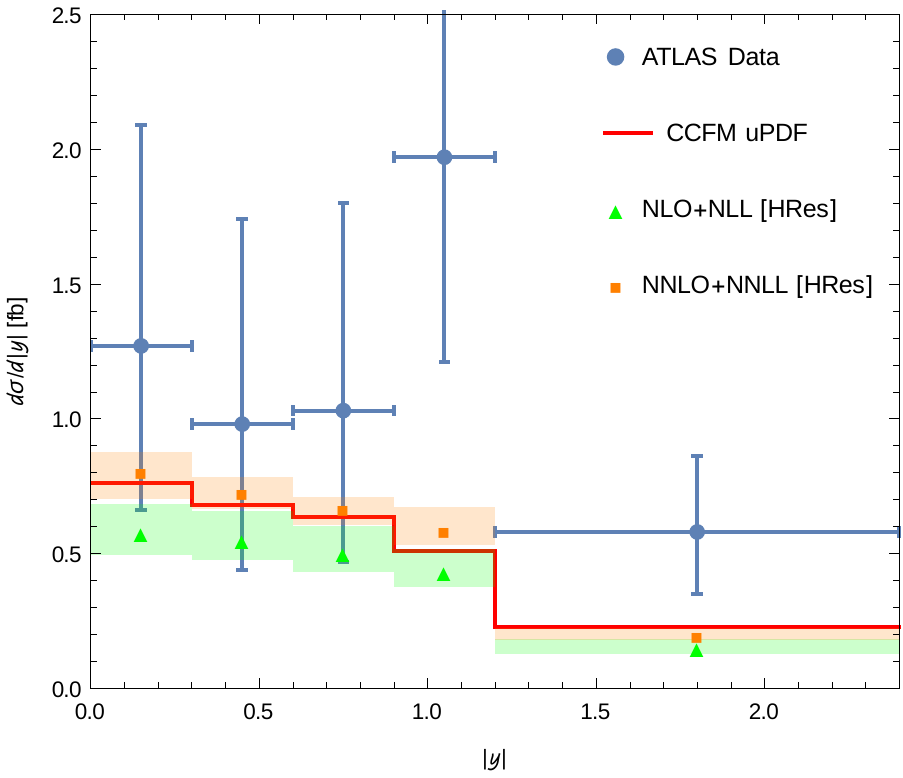}\label{fig1b}}
 \caption{Differential cross section of the Higgs boson production in four-lepton
          decay channel at $\sqrt{s}$ = 8 TeV. (a) and (b) are the differential
          distributions in transverse momentum of the four-leptons ($p_T$) and
          rapidity ($y$) of the four-leptons respectively. Solid line (red) is
          a results obtained using $k_T$-factorization approach with CCFM gluon uPDFs.
          Filled triangle and filled square points corresponds to estimated
          obtained using HRes tool up to NNLO+NNLL accuracy and shaded region
          corresponds to scale uncertainty in renormalization and factorization
          scale. Experimental data points are from ATLAS. The error bars on the
          data points shows total (statistical $\oplus$ systematic) uncertainty.}
\label{fig1}
\end{figure}
\begin{figure}[!ht]
  \centering
  \subfloat[]{\includegraphics[trim=0 0 0 0,clip,height=0.44\linewidth,width=0.49\linewidth]{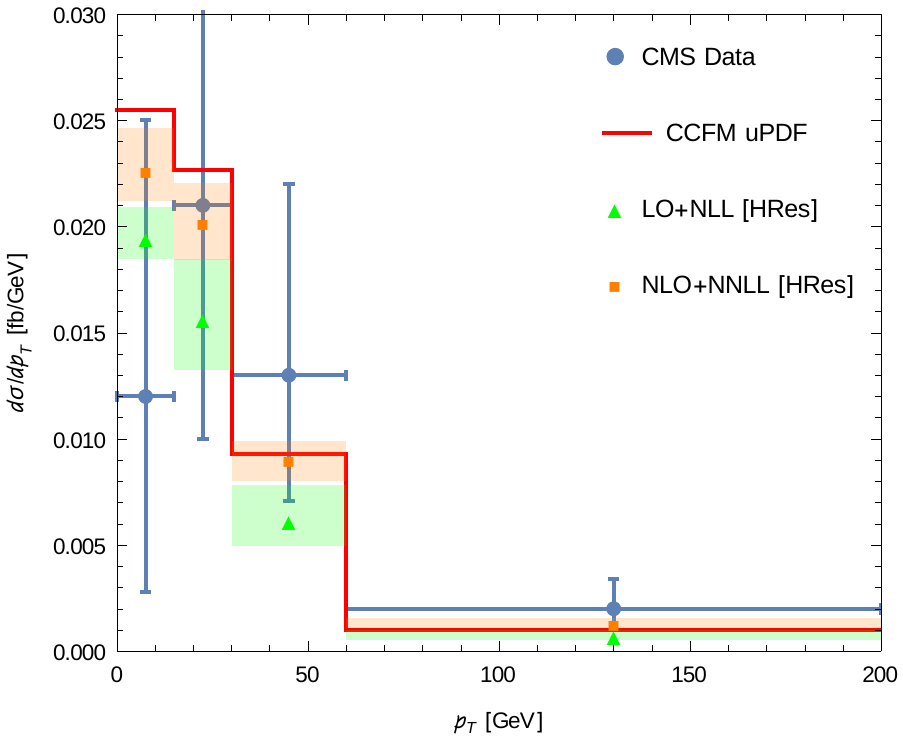}\label{fig2a}}
  \subfloat[]{\includegraphics[trim=0 0 0 0,clip,height=0.44\linewidth,width=0.49\linewidth]{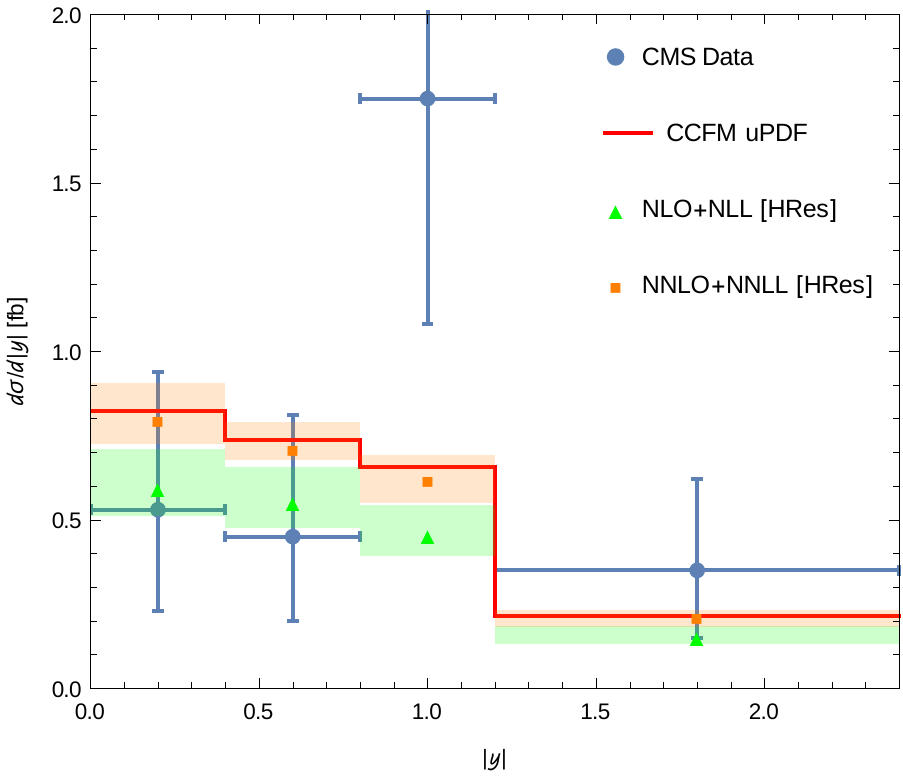}\label{fig2b}}
 \caption{Differential cross section of the Higgs boson production in four-lepton
          decay channel at $\sqrt{s}$ = 8 TeV. (a) and (b) are the plots of
          differential distributions in transverse momentum of the four-leptons
          ($p_T$) and rapidity ($y$) of the four-leptons respectively. Solid line
          (red) is a results obtained using $k_T$-factorization approach with CCFM
          gluon uPDFs. Filled triangle and filled square points corresponds to
          estimated obtained using HRes tool up to NNLO+NNLL accuracy and shaded
          region corresponds to scale uncertainty in renormalization and factorization
          scale. Experimental data points are from CMS.  The error bars on the data
          points shows total (statistical $\oplus$ systematic) uncertainty}  
\label{fig2}
\end{figure}

\section{Numerical Estimates}

We have calculated differential cross section for the Higgs boson decay into four leptons 
at $\sqrt{s} = 8$~TeV. Total decay width and mass of the Higgs boson is set
to be equal to 4.0 MeV and 125.09~GeV respectively \cite{Agashe:2014kda}. Our results
are compared with recent experimental data from ATLAS and CMS collaborations~\cite{
Aad:2014tca,Khachatryan:2015yvw}. We have also compared our estimates with a results
obtained using HRes tool \cite{deFlorian:2012mx,*Grazzini:2013mca} which gives a
fixed order cross section calculated up to next to next to leading order plus next
to next leading logarithm (NNLO+NNLL) accuracy within collinear factorization formalism.

In Figure \ref{fig1}, we have given an estimates obtained using $k_T$-factorization approach
to compare with the ATLAS measurements \cite{Aad:2014tca}. Our results obtained using 
$k_T$ factorization are close to NNLO+NNLL results obtained using HRes tool. 
We have done a comparison for both $p_T$ and $y$ differential distribution. 
In Figure \ref{fig2}, our results are compared for CMS experiment
\cite{Khachatryan:2015yvw}. We see a similar behaviour in the comparison with the data
from the CMS experiment as well. Our results are consistently close to NNLO+NNLL results
which are fixed order calculation within collinear factorization approach. This can
be explained considering the fact that the main part of higher order corrections
included in the $k_T$-factorization approach~\cite{Lipatov:2005at,*Ryskin:1999yq}.

\section{Conclusion}

Here we have presented our results of the differential cross section for the Higgs boson 
production within $k_T$-factorization approach. We have compared our results with 
the fixed order estimates up to the order NNLO+NNLL within collinear factorization obtained using  HRes tool.
We have plotted our results against recent experimental measurements of the differential 
cross section for the Higgs boson production in four-lepton decay channel from the ATLAS and CMS Collaborations. 
In summary, we have done a phenomenological study for the case of the Higgs boson
production in the four-lepton decay channel considering the framework of $k_T$-factorization.
In this phenomenological study we have used CCFM unintegrated parton densities.


\section*{Acknowledgements}
Authors would like to extend sincere gratitude to V. Ravindran for continuous support and fruitful discussions during the course of this work and A. Lipatov for providing their codes of a similar study for diphoton distributions.
VR would like to thank Giancarlo Ferrera for useful discussions on HRes tool and related topics. 
RI would also like to acknowledge the hospitality provided by the Institute of Mathematical Sciences, 
Chennai, India where a part of work has been done.



\bibliography{TMD}
\bibliographystyle{JHEP}

\end{document}